\documentclass[aps,pra,twocolumn,showpacs,superscriptaddress,groupedaddress,amsmath,amssymb]{revtex4-1}

\usepackage{xparse} 
\usepackage{graphicx}
\usepackage{amsmath}
\usepackage{color}
\usepackage{dsfont}
\usepackage{hyperref}
\usepackage[margin=50pt]{geometry}
\usepackage[caption=false]{subfig}

\makeatletter
\newcommand*{\balancecolsandclearpage}{%
  \close@column@grid
  \clearpage
  \twocolumngrid
}
\makeatother

\begin{document}

\title{The impact of scaffolding and question structure on the gender gap}
\author{Hillary Dawkins, Holly Hedgeland, and Sally Jordan}
\affiliation{
School of Physical Sciences, The Open University, Walton Hall, Milton Keynes, MK7 6AA, UK
}
\date{\today}

\begin{abstract}

We address previous hypotheses about possible factors influencing the gender gap in attainment in physics. 
Specifically, previous studies claim that male advantage may arise from multiple-choice style questions, and that scaffolding may preferentially benefit female students. We claim that female students are not disadvantaged by multiple-choice style questions, and also present some alternative conclusions surrounding the scaffolding hypothesis. By taking both student attainment level and the degree of question scaffolding into account, we identify questions which exhibit real bias in favour of male students. We find that both multi-dimensional context and use of diagrams are common elements of such questions.  

\end{abstract}

\maketitle

\section{Introduction} 

The gender gap in attainment in physics is consistent and well documented. Across institutions, male students outperform their female counterparts in terms of undergraduate course performance \cite{Dockter:2008, Wee:1993, Andersson:2016}, as well as outcome on subject specific concept inventories (FCI \cite{Dockter:2008, Bates:2013, Brewe:2010, Richardson:2013, Lorenzo:2006}, BEMA \cite{Lauer:2013, Kost:2010, Pollock:2008}, and CSEM \cite{Pollock:2008,Kohl:2009}). At the UK Open University (OU), we observe a significant difference in attainment on the second level physics modules in favour of males, and furthermore this gap is persistent across multiple years of instruction.   

While the existence of a real and significant gap is well established, the contributing factors are less well understood (see \cite{Madsen:2013} for a review of 17 studies). Possible factors include background and preparation, of which many possible measures exist. Previous studies identify concept inventory pretest scores \cite{Kost:2009,Kost:2010}, SAT math scores \cite{Kost:2009,Kost:2010,Lorenzo:2006}, ACT math scores \cite{Kost:2009,Kost:2010}, and prerequisite course grades \cite{Kost:2010} to vary significantly by gender. Sociocultural factors may also play a role, for example self-efficacy and CLASS scores \cite{Adams:2006} (a measure of learning attitudes about science). Finally, there is the issue of question construction including type of question (constructed response, multiple choice, or other selected response), presentation (graphs, diagrams, words), and male-biased context (references to sports and cannons). Here we focus on identifying factors from the final category of question structure, as these are the most readily modified.       

A recent study from the University of Cambridge \cite{Gibson:2015} observes an interesting dependence on question structure in the form of scaffolding. Scaffolding refers to the degree to which a question guides the student through the problem-solving process. Previous studies support the use of scaffolding in aiding students' learning and conceptual understanding in physics \cite{Ding:2011, Lin:2015, Lindstrom:2011}. However, \cite{Gibson:2015} is the first study to our knowledge to identify a dependence on gender. It is therefore important to verify these findings across institutions and student populations, prior to taking action towards any instructional reform.     

In light of the large and diverse student population of the Open University, we find ourselves well situated to address these issues. The goals of the present study are to
\begin{enumerate}
\item Identify elements of question structure which may be disadvantaging female students
\item Test the scaffolding hypothesis as a potential solution
\end{enumerate}
Taking student ability (as measured by overall attainment levels) and question difficulty into account, we identify questions that pose significant male bias and those which do not. We discuss our findings in the context of current literature on the subject. Furthermore we challenge the conclusions presented in \cite{Gibson:2015}, and offer some alternative conclusions.

\section{Context}

The present study examines gender differences in attainment observed in the second level (FHEQ Level 5) physics modules at the Open University. We first spend some time reviewing the structure of the Open University, the modules in question, and the student population.  

The Open University approaches higher education in a non-traditional way in that there are no admission requirements, and modules are completed at a distance with substantial online elements. Students select and complete modules, according to their needs, to make up a degree comprised of 360 credits if desired. Students are attracted to the open concept for a variety of reasons including flexibility, part-time options, returning to study later in life, and completing second degrees. We therefore expect that the student population is demographically diverse. Despite differences in the student population, similar trends in attainment gaps have been identified as at other institutions. Of particular interest is a large gap in attainment at the second level, the first level at which physics is taught as a separate module, which does not exist at lower or higher levels.

The 60-credit second level physics modules (previously S207, now S217) include mechanics, thermodynamics, electricity and magnetism, quantum physics, and nuclear physics at an introductory to intermediate level. Although prerequisites are not enforced, it is expected that students will have completed the introductory level one science module, from which they will have gained some familiarity with some of these topics as well as appropriate mathematical preparation. The module population comprises a mixture of students intending to take further physics modules and those intending to take further science or mathematics modules outside of physics. 

Throughout the module, students complete interactive computer-marked assignments (iCMAs), which are short problems requiring numeric open responses or selected responses, in addition to tutor-marked assignments. Students receive feedback on their iCMA answers and are permitted to retry questions as many times as desired. The module ends with an exam which contains, among other components, long answer open response questions. In this study, we analyze iCMA questions to identify any gender bias, and look at exam long answer questions to address the scaffolding hypothesis. Data was collected over four recent presentations of the module; S207 in 2012-2014 and S217 in 2015. The total number of students completing the module in this time period was 5535, 4286 (77\%) males and 1249 (23\%) females.   

\section{Identifying Bias}
\subsection{The Mantel-Haenszel method}

The Mantel-Haenszel method is a statistical technique used to identify differences between groups using a stratified data set \cite{Osterlind:2009}. The idea is that possible confounding variables will be captured by the stratification. 

In this case, we wish to detect iCMA questions which exhibit significant male bias while accounting for student ability and question difficulty. Therefore we take our two groups to be male and female students, and students are stratified according to ability as measured by their overall performance on iCMA questions. 
Table \ref{tab:MH_cross} shows a cross-table representing the number of students in each group answering an item correctly at the $i$th stratum.  
\begin{table}
\begin{ruledtabular}
\caption{\label{tab:MH_cross}A cross-table of the $i$th stratum depicting the number of students in each group (male and female) to get a particular iCMA question correct or incorrect on the first attempt. The total number of students in the $i$th stratum is $N_i = m_i^1 + m_i^0 + f_i^1 + f_i^0$.}
\begin{tabular}{ccc} 
 & Correct (1)  & Incorrect (0) \\ \hline
Male & $m_i^1$ & $m_i^0$\\
Female & $f_i^1$ & $f_i^0$\\
\end{tabular}
\end{ruledtabular}
\end{table}
For each item, we calculate the odds ratio (ratio of success probabilities between groups) of the $i$th stratum as $m^1_if^0_i/m^0_if^1_i$. A weighted average across all strata then provides the overall odds ratio for a particular question, referred to as the Mantel-Haenszel alpha:
\begin{align}
\alpha_{MH} = \frac{\sum_im^1_if^0_i/N_i}{\sum_im^0_if^1_i/N_i}.
\end{align} 
For ease of comparison, this is often converted to a logarithmic scale as
\begin{align}
\alpha^*_{MH} = -2.35 \ln \Bigg( \frac{\sum_im^1_if^0_i/N_i}{\sum_im^0_if^1_i/N_i} \Bigg).
\end{align}
The sign and magnitude of $\alpha^*_{MH}$ indicate the direction and strength of bias within a question. Negative values indicate a bias in favour of males, meaning that male students have a greater probability of answering this question correctly compared to female students of equal ability. Likewise, positive values indicate a bias in favour of females. The absolute value of $\alpha^*_{MH}$ indicates the strength of the bias, and is deemed to be significant if $|\alpha^*_{MH}| \geq 1$ \cite{Zwick:2012}.  

As a second assurance of significance, each $\alpha_{MH}$ value is tested using a chi-squared distribution. In this case, the null hypothesis is that the odds ratio is equal to one at each stratum, and the alternative hypothesis is that at least one odds ratio is
different from unity \cite{Osterlind:2009}. In this study, we flag questions as having significant bias if both conditions i) $|\alpha^*_{MH}| \geq 1$ and ii) $p \leq .05$ are satisfied.    

\subsection{Analysis and results}

Applying the Mantel-Haenszel method to 56 iCMA questions flags 3 questions of significant bias, all in favour of male students. Further, 2 questions were noted to be of interest having significant $p$-values and insignificant $|\alpha^*_{MH}|$ values, but in favour of female students. These items were included for being the only questions of some significance with female bias. Table \ref{tab:iCMAlevels} shows the $\alpha^*_{MH}$ values with significance levels for each question of interest. 
\begin{table}
\begin{ruledtabular}
\caption{\label{tab:iCMAlevels}Strength of bias ($\alpha_{MH}^*$) and significance ($p$) values for iCMA questions of interest.}
\begin{tabular}{cccccc} 
 & $M_1$ & $M_2$ & $M_3$ & $F_1$  & $F_2$ \\ \hline
$\alpha_{MH}^*$ & -2.9 & -3.2 & -3.3 & .14 & .39 \\
$p$ & .045 & .013 & .032 & .016 & .037 \\
\end{tabular}
\end{ruledtabular}
\end{table}
Questions are labeled as $M_1,M_2,M_3$ (those having male advantage) and $F_1,F_2$ (those having female advantage). We note that $M_1$, $M_2$, and $M_3$ all display very strong levels of bias with $|\alpha^*_{MH}|$ values well above the threshold. 

\begin{figure*}

\begin{minipage}{.5\linewidth}
\centering
\subfloat[$M_1$: Inclined plane]{\label{main:a}\includegraphics[scale=.7]{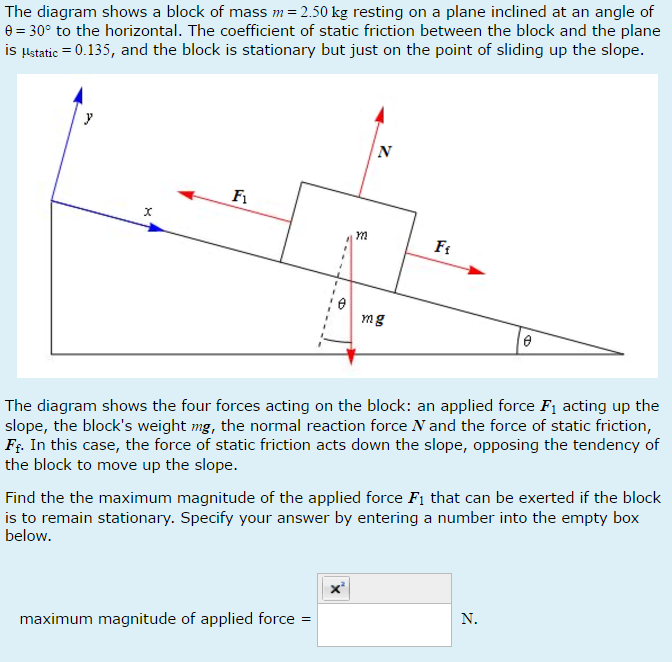}} \\ \vfill
\subfloat[$M_2$: Torque]{\label{main:c}\includegraphics[scale=.7]{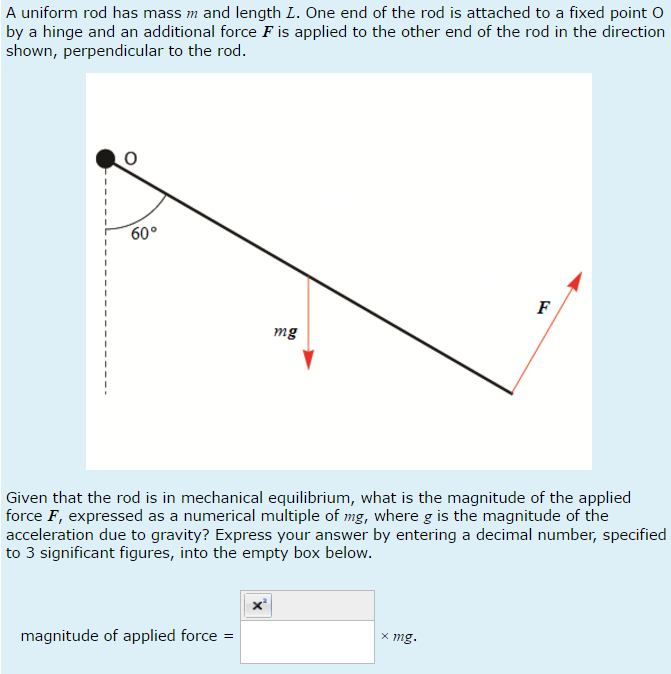}}
\end{minipage}%
\hfill
\begin{minipage}{.5\linewidth}
\centering
\subfloat[$M_3$: Generic PVT surface]{\label{main:b}\includegraphics[scale=1.03]{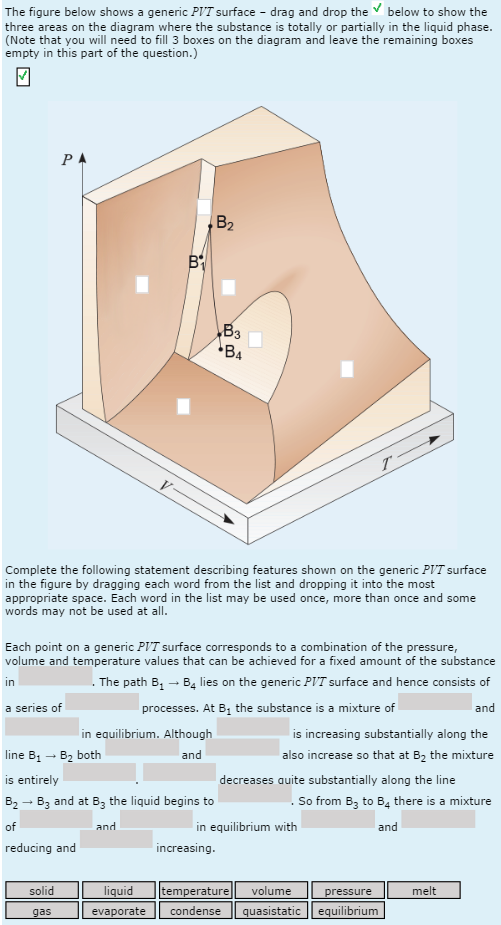}}
\end{minipage}\par\medskip

\caption{iCMA items displaying bias in favour of male students}
\label{fig:iCMA_Mquestions}
\end{figure*}

Fig.~\ref{fig:iCMA_Mquestions} shows the items displaying male bias. 
Notably, all questions require interpreting a diagram of more than one dimension, which we find to be consistent with current literature. Wilson \textit{et al.} \cite{Wilson:2016} studied the impact of question structure on the gender gap along five broad dimensions: content, process required, difficulty, presentation and context. They observed large gender gaps in favour of males for questions which involved the process of interpreting a diagram, which presented the question using a significant diagram, and which involved more than one spatial dimension. Studies which aim to identify gender gaps on FCI questions have observed the largest disparities on items 6 (path of ball leaving a channel), 12 (path of cannonball fired off a cliff) \cite{Dietz:2012}, 14 (path of object released from an airplane), and 23 (path of a rocket after thrust is turned off) \cite{Dockter:2008}. Clearly all of these items involve predicting motion in two dimensions, and all are presented using a diagram. The observed gender gap on projectile-motion-like items is sometimes ascribed to male-biased context \cite{Rennie:1998}. However, attempts to reword FCI items in a more traditionally female context have failed to improve female performance \cite{McCullough:2004}. In light of this discussion we find the inclusion of item $M_3$ particularly interesting. The content is thermodynamics, far removed from kinematics or predicting motion. The context is certainly not experienced or male-biased, and yet a large and significant gap is observed. The only identifiable common trait among all items is the need to interpret a multi-dimensional diagram. 

\begin{figure*}
\subfloat[$F_1$: Quantum physics]{%
  \includegraphics[scale = .7]{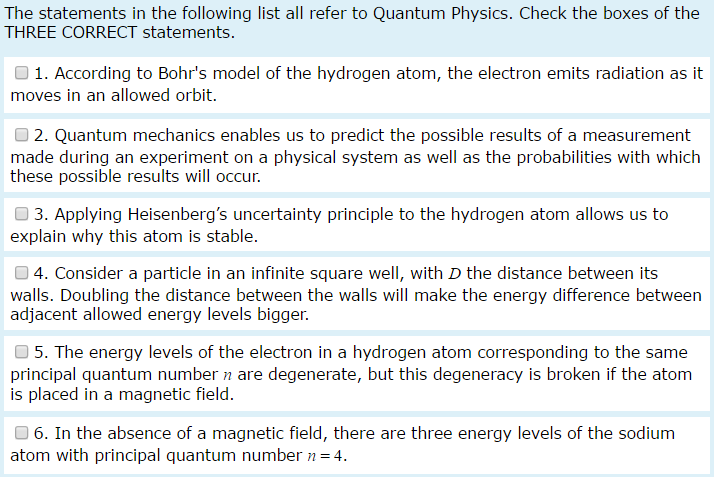}%
}\hfill
\subfloat[$F_2$: Predicting motion]{%
  \includegraphics[scale = .7]{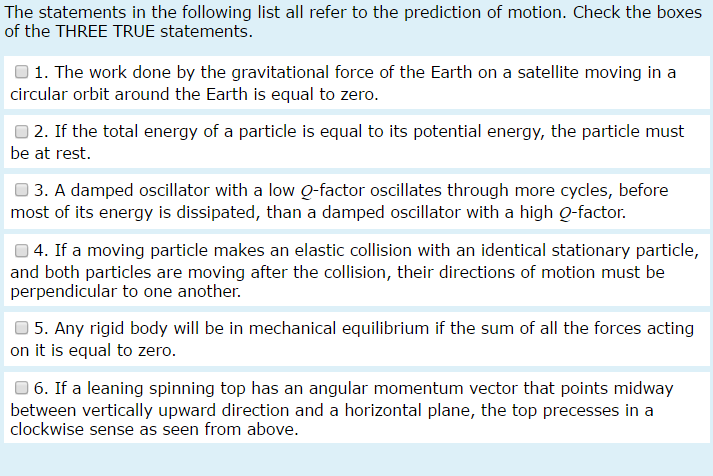}%
}
\caption{iCMA questions displaying bias in favour of female students}
\label{fig:iCMA_Fquestions}
\end{figure*}

Fig.~\ref{fig:iCMA_Fquestions} shows the items displaying female bias. 
As previously stated, these questions have significant $p$-values but do not have significant $|\alpha^*_{MH}|$ values, implying that the bias is small. Nonetheless, these items are of interest as the only female-biased questions of some significance. Both items involve careful reading, a task suggested to have a female advantage \cite{McBride:2009}. Interestingly, item $F_2$ is on the subject of predicting motion.
This observation further supports the idea that male bias arises from the need to interpret a diagram or multi-dimensional context, rather than content related to predicting motion. 

Other important observations arise from those questions which were \textit{not} deemed to have significant bias. In particular, we address the widely held belief of male advantage on multiple choice style questions \cite{Shakhar:1991, Everaert:2012, Hudson:2010, Hamilton:1998}. Of 20 iCMA questions presented in multiple choice format, none are observed to have a significant gender gap. These include questions similar in content to items $M_1$, $M_2$, and $M_3$. Furthermore when item gaps are ranked in order of significance, we find that multiple choice questions populate the side of the spectrum of lesser significance. We conclude that there is no evidence to suggest a female disadvantage owing to multiple choice structured questions. 

\section{Scaffolding}

\subsection{Scaffolding definition}

Scaffolding is broadly defined to have occurred when an expert or more knowledgeable person helps a learner to accomplish tasks that would otherwise be unattainable \cite{Wood:1976}. A traditional example would be a teacher providing strategic guidance and feedback while a student completes a problem. In more recent years this definition has evolved to include interactive computer-assisted learning, as well as peer instruction and similar socialized learning environments \cite{Lin:2012}. 

Due to widespread usage of the term ``scaffolding" in multiple circumstances, it is important to carefully define the term in the context of physics education research. In the present study, we consider scaffolding only as it may be applied to written exam questions. We define 6 general ways in which scaffolding can occur (elements), and further provide specific instances of each that are likely to be encountered in physics problems. Table \ref{tab:scaff_list} shows a complete itemization of the elements. 
\begin{table*}
\begin{ruledtabular}
\caption{\label{tab:scaff_list}The 6 elements of scaffolding (bold), with itemized examples of how each element is likely to appear in written physics problems.}
\begin{tabular}{l} 
\textbf{use of representations and language to bridge expert-novice understanding}\\ \hline
1. technical words are described in everyday language\\
2. mathematical symbols are explained in words\\
3. a diagram is used to give meaning to technical words or symbols\\ \hline

\textbf{reduction of cognitive overhead}\\ \hline
4. includes a math (or other background) reminder\\
5. somehow automates a routine task (eg. unit conversions given, constants given that could have been looked up)\\
6. no penalty for missing sig figs, wrong unit, wrong numeric value or other nonsalient component of the question\\
7. provides a diagram or graph that the student could have constructed with the available information\\ \hline

\textbf{insertion of expert knowledge}\\ \hline
8. expert directed focus is used (eg. key information is highlighted using bold or italicized text)\\
9. explicitly instructs student to make an expert assumption (eg. ``you may ignore air resistance")\\
10. the student is warned of a common mistake or relevant misconception\\ \hline

\textbf{ordered task decomposition (provide structure for complex tasks)}\\ \hline
11. each part of the question contains only one expected output (numeric or otherwise)\\
12. an output (numeric or otherwise) is required in subsequent work\\
13. marks are awarded for interpreting outputs (no further calculation required) \\ 
14. question has a wide mark distribution (each part is worth less than 50\% of the total awarded marks)\\ \hline

\textbf{conceptual prompting}\\ \hline
15. asks student to define or explain an equation that they should use \\
16. asks student to identify a concept that they should make use of \\
17. asks student to draw a diagram before beginning the problem \\ \hline

\textbf{reduction of degrees of freedom}\\ \hline
18. gives student the appropriate equation to use\\
19. prompts at how the question is expected to be solved (eg. ``using the principle of conversation of energy...")\\
20. explicitly instructs student on how to begin a task\\
\end{tabular}
\end{ruledtabular}
\end{table*}
Many elements are adapted from the guidelines outlined in \cite{Quintana:2004}, which combines theoretical foundations with prior work to define a common framework for scaffolding within computer-assisted assignments. The element of conceptual prompting is motivated by \cite{Ding:2011}. There it was shown that students will successfully apply physics concepts to problems if they are prompted to identify the concept immediately beforehand. Taken together, the elements listed in Table \ref{tab:scaff_list} define what is meant by scaffolding within this study.    


\subsection{Gains by gender}

In a study on question structure and its impact on the gender gap, Gibson \textit{et al.} \cite{Gibson:2015} administered 2 versions of an exam. One exam used highly scaffolded questions, and one used traditional exam style questions. Between the low and high scaffolding versions, female students achieved a gain in exam score of 13.4\% while male students achieved a gain of 9.0\%. The study therefore concludes that scaffolding benefits all students, but that female students benefit preferentially. We observe no such preferential treatment, and argue that other factors may be at play.  

Using the elements of scaffolding and individual items as a scoring system, all exam questions were assigned a ``scaffolding score". Questions displaying 2 or fewer items were labeled as low scaffolding, and questions with 7 or more items were labeled as high scaffolding. All questions belonging to either group can be found in the appendix. Fig. ~\ref{fig:exam_scores} shows the performance of students on each question by gender, and Table \ref{tab:exam_table} shows the average performance as well as gains provided by increased scaffolding. The average gain is 6.6\% for female students, and 5.2\% for male students. 
\begin{figure*}
\subfloat[Low scaffolding questions]{%
  \includegraphics[scale = 1.06]{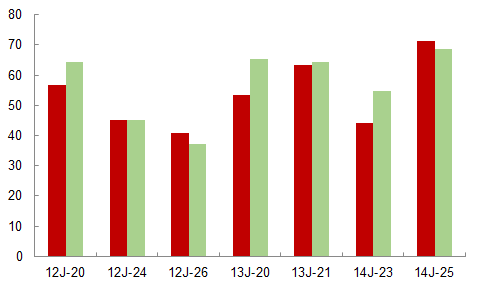}%
}
\subfloat[High scaffolding questions]{%
  \includegraphics[scale = 1.06]{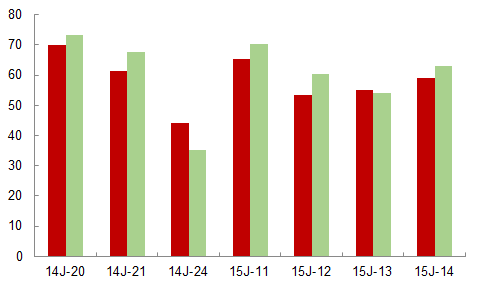}%
}
\caption{\label{fig:exam_scores} Average performance (percentage score) of female (red) and male (green) students on all exam questions belonging to the low (a) and high (b) scaffolding groups. Exam questions are labeled as they appear in the appendix.}
\end{figure*}
\begin{table}
\begin{ruledtabular}
\caption{\label{tab:exam_table} Average performance by gender on exam questions assigned to the low and high scaffolding groups. Difference represents the average difference in exam grade between genders. Gain represents the increase in average exam score as a result of increased scaffolding.}
\begin{tabular}{c|cc|c} 
 & Male & Female~~~~~ & Difference~~~~~  \\ \hline
Low scaffolding & 65.4& 62.5& 2.9 \\
High scaffolding& 70.5& 69.1& 1.4\\ \hline
Gain & 5.2&6.6 & \\
\end{tabular}
\end{ruledtabular}
\end{table}

Although not as clear, the data does at first glance seem to support the conclusions of \cite{Gibson:2015}. Male students outperform female students on the low scaffolding questions by 2.9\% ($p = .087$), and by only 1.4\% on the high scaffolding questions ($p = .42$). However neither result is statistically significant, and we should also consider how scaffolding benefits students performing at different levels. Intuitively, we expect that scaffolding cannot greatly benefit the highest achieving students (who likely know the information and do not have much room to improve) or the lowest achieving students (who are too unprepared for scaffolding to provide a use). Students completing module S207 and S217 are assigned a level (1-4) based on overall performance on the module (1 being the highest level of achievement). Table \ref{tab:levels} shows the average score of students on the low and high scaffolding questions by level, as well as the number of male and female students in each level. 
\begin{table}
\begin{ruledtabular}
\caption{\label{tab:levels} Average performance by level on exam questions assigned to the low and high scaffolding groups. Gain represents the increase in average exam score as a result of increased scaffolding. $N$ represents the total number of students achieving each level by gender.}
\begin{tabular}{ccccc} 
 Level  &  1&2&3&4  \\ \hline
Low average & 90.5 & 69.5 & 52.5 & 40.7 \\
High average & 90.8 & 76.5 & 61.2 & 46.1 \\
Gain& 0.33 & 7.0 & 8.7 & 5.4 \\
$N$ males & 414 & 732 & 577 & 297 \\
$N$ females & 85 & 248 & 151 & 71 \\
\end{tabular}
\end{ruledtabular}
\end{table}
As expected, scaffolding provides the greatest gains to the intermediate students. Performing a weighted average of gains across level by the number of female and male students in each level can give us an idea of the expected gains by gender. Doing this, we estimate expected gains of 6.2\% for female students, and 5.8\% for male students. The expected gain is higher for female students as a consequence of the fact that fewer female students achieve a level 1. The expected gains are not significantly different than the actual gains for either gender, and therefore we conclude that preferential female gain is simply an artifact of gain dependency on level.     

\subsection{Questions of interest}

\begin{figure}
\subfloat[$L$: Low scaffolding]{%
  \includegraphics[scale = .45]{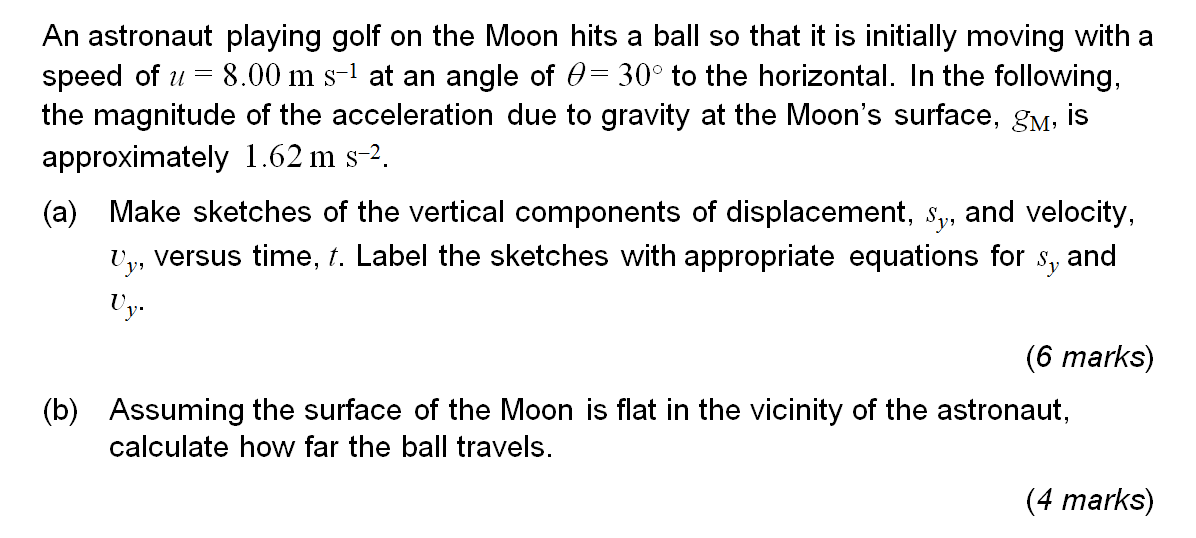}%
}\\
\subfloat[$H$: High scaffolding]{%
  \includegraphics[scale = .57, angle = 90]{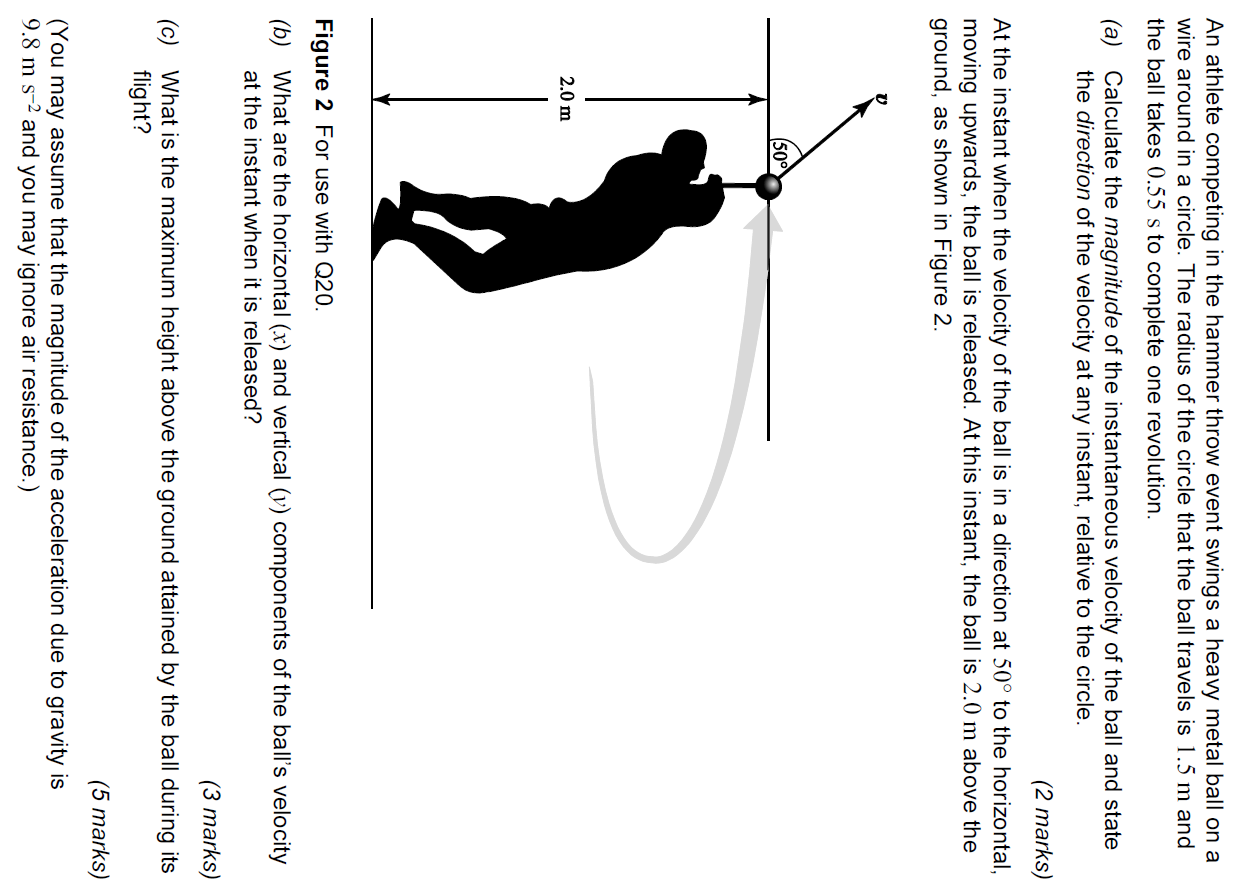}%
}
\caption{A pair of 2-dimensional projectile motion problems, of different scaffolding levels. Male students very significantly outperform female students on $L$, but performance is equal across gender on $H$.}
\label{fig:exam_interest}
\end{figure}

Although scaffolding does not appear to preferentially benefit female students in general, we note some particular questions of interest. Fig.~\ref{fig:exam_interest} shows one question from the low scaffolding group ($L$), and one question from the high scaffolding group ($H$).
Both are 2-dimensional projectile motion questions, but display significant performance differences. Table \ref{tab:interest} shows the average performance on each question, and the difference between genders with significance levels. 
\begin{table}
\begin{ruledtabular}
\caption{\label{tab:interest} Average performance of students on questions of interest $L$ and $H$ by gender. The difference between male and female attainment is displayed along with the significance level ($p$-values).}
\begin{tabular}{ccccc} 
 &  Male & Female & Difference&Significance ($p$) \\ \hline
$L$& 64.5 & 56.6 & 7.8 & 0.013 \\
$H$ & 73.3 & 70.0 & 3.3 & 0.46 \\
\end{tabular}
\end{ruledtabular}
\end{table}
Of all exam questions, $L$ exhibits one of the most significant differences in performance between genders, and $H$ shows no significant difference. The scaffolding gains are comparable to those observed in \cite{Gibson:2015} (13.4\% for females, 8.8\% for males). We conclude that scaffolding may play a role in reducing the gender gap in specific types of problems which were previously identified to contain a male bias, namely questions involving multi-dimensional context. 

\section{Discussion and conclusions}

In summary, we have identified elements of question structure that promote male bias, and further address the scaffolding hypothesis as a potential solution. We conclude that neither the use of multiple-choice style questions nor the level of scaffolding can sufficiently explain the gender gap.

We have used a Mantel-Haenszel stratified analysis to account for student ability, and find iCMA questions with significant performance differences between genders. By flagging only those questions which display significant bias in both measures ($|\alpha_{MH}^*|$ and $p$), we have reduced the possibility of flagging false positives. We therefore conclude that the 3 flagged questions exhibit real and significant male bias. All questions involve interpreting a diagram, and all involve multi-dimensional context. Our findings are in agreement with \cite{Wilson:2016}, and similar studies on the FCI \cite{Dockter:2008, Dietz:2012}. Because multi-dimensional diagrams appear most frequently in mechanics problems, previous studies may have incorrectly attributed male bias to mechanics content. Further investigation with more types of questions will be required to separate the variables of content and presentation. 

Scaffolding has recently been argued to preferentially benefit female students \cite{Gibson:2015}, and therefore have the potential to aid in reducing the gender gap. The study of Gibson \textit{et al.} uses a smaller number of students and less varied exam content than the present study to reach this conclusion. In a similar analysis, we do not observe a dependence on gender, and argue that any perceived dependence is actually due to student achievement level. The advantage of \cite{Gibson:2015} is that exam questions were designed specifically to measure scaffolding gains, whereas the present study collected data from actual exam responses. Therefore questions between the low and high scaffolding groups do not match onto each other exactly as in \cite{Gibson:2015}. Future studies can make use of the elements of scaffolding to produce low and high scaffolding versions of the same question for use in experimental exams. 

Even if scaffolding does not preferentially benefit female students in general, it may still play a role in reducing the gender gap. We make note of a pair of questions involving multi-dimensional context (2D projectile motion), for which the gap is reduced between low and high scaffolding versions. If male bias within a question can be reduced by increased scaffolding for novice students, then this provides a route to addressing gender gaps in attainment.  

\section{Acknowledgments}

The authors acknowledge financial support from eSTEeM, the OU centre for STEM pedagogy, as well as the co-operation of the S207 and S217 module teams, and useful discussions with Richard Jordan.

\balancecolsandclearpage
\appendix*

\section{Exam questions labeled as low and high scaffolding}\label{app}

\begin{figure}[h]
\includegraphics[scale=.63]{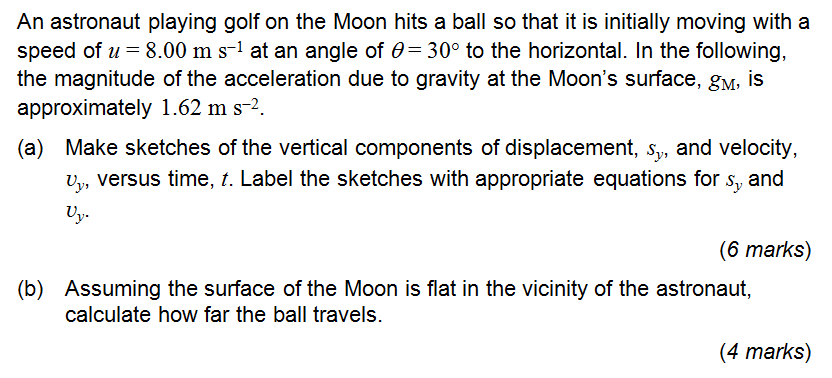}
\caption{\label{fig:12J20} Low scaffolding: 12J-20}
\end{figure}

\begin{figure}[h]
\includegraphics[scale=.63]{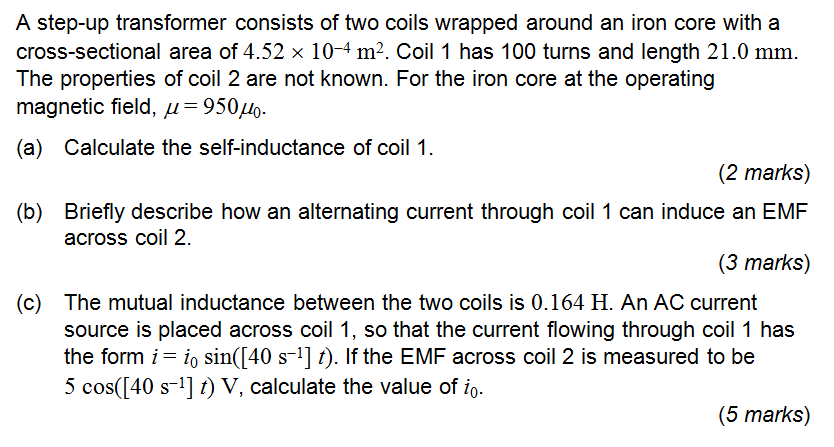}
\caption{\label{fig:12J24} Low scaffolding: 12J-24}
\end{figure}

\begin{figure}[h]
\includegraphics[scale=.63]{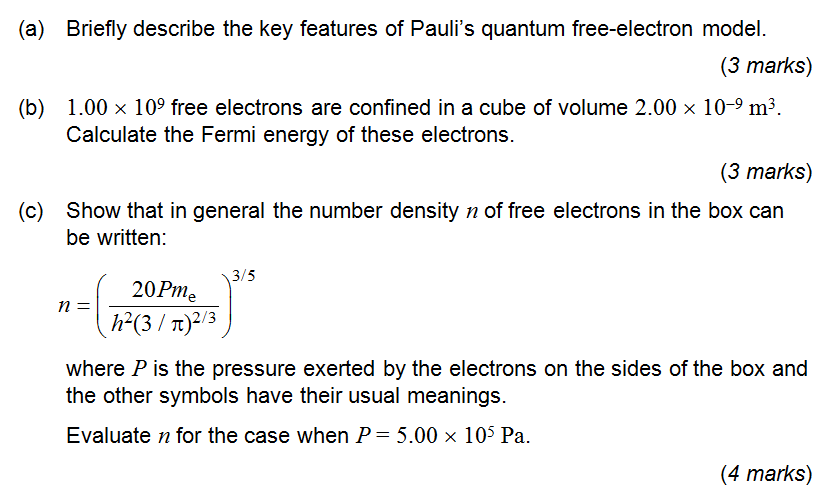}
\caption{\label{fig:12J26} Low scaffolding: 12J-26}
\end{figure}

\begin{figure}[h]
\includegraphics[scale=.6, angle = 90]{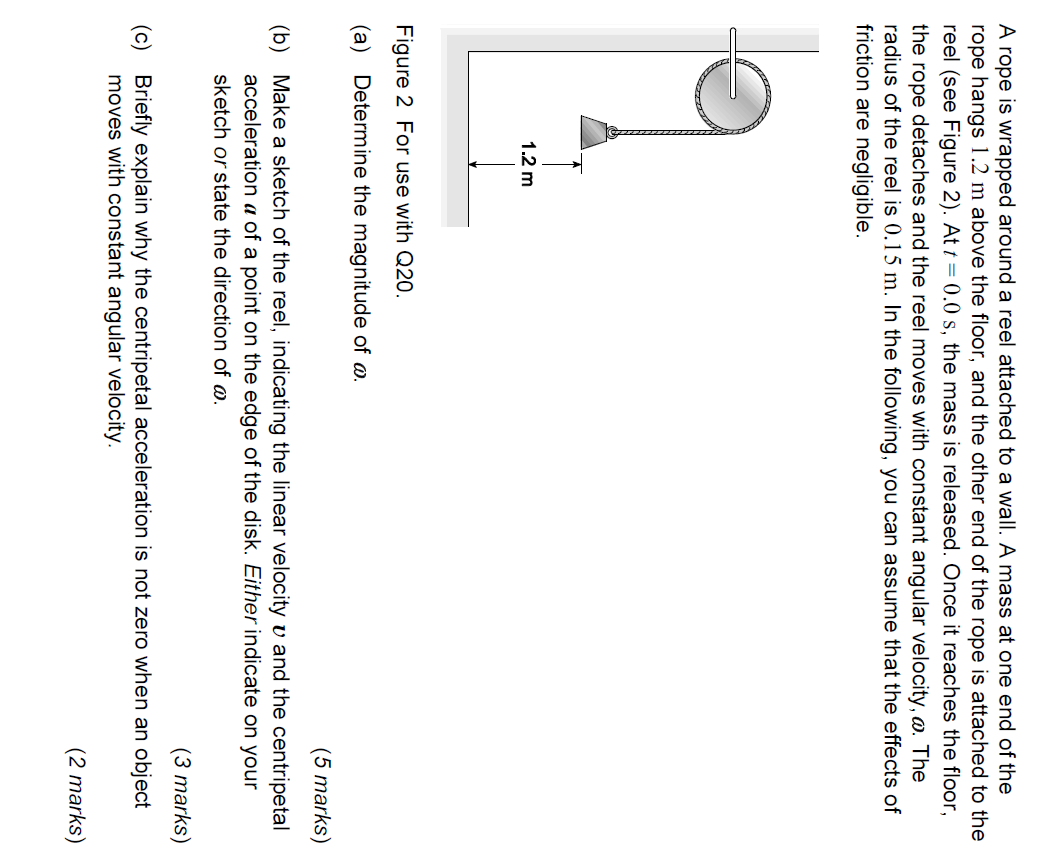}
\caption{\label{fig:13J20} Low scaffolding: 13J-20}
\end{figure}

\begin{figure}[h]
\includegraphics[scale=.63]{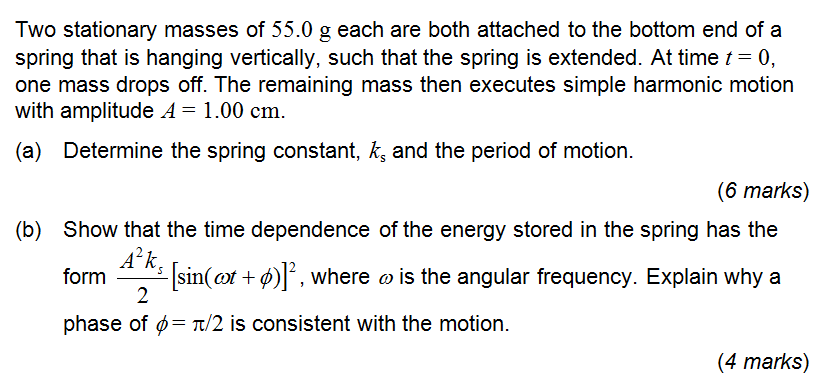}
\caption{\label{fig:13J21} Low scaffolding: 13J-21}
\end{figure}

\begin{figure}[h]
\includegraphics[scale=.63]{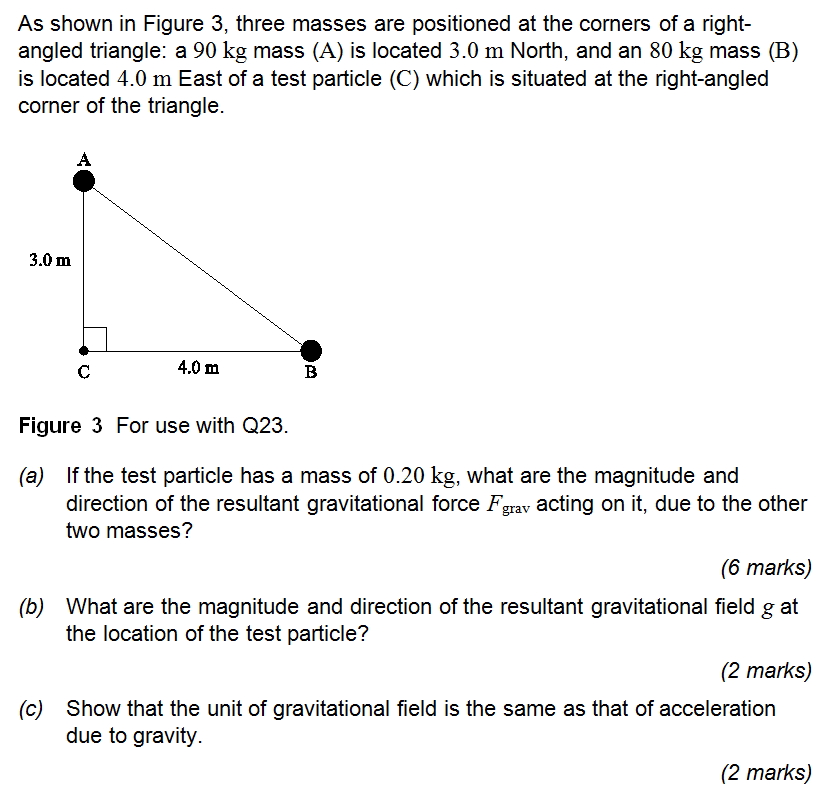}
\caption{\label{fig:14J23} Low scaffolding: 14J-23}
\end{figure}

\begin{figure}[h]
\includegraphics[scale=.63]{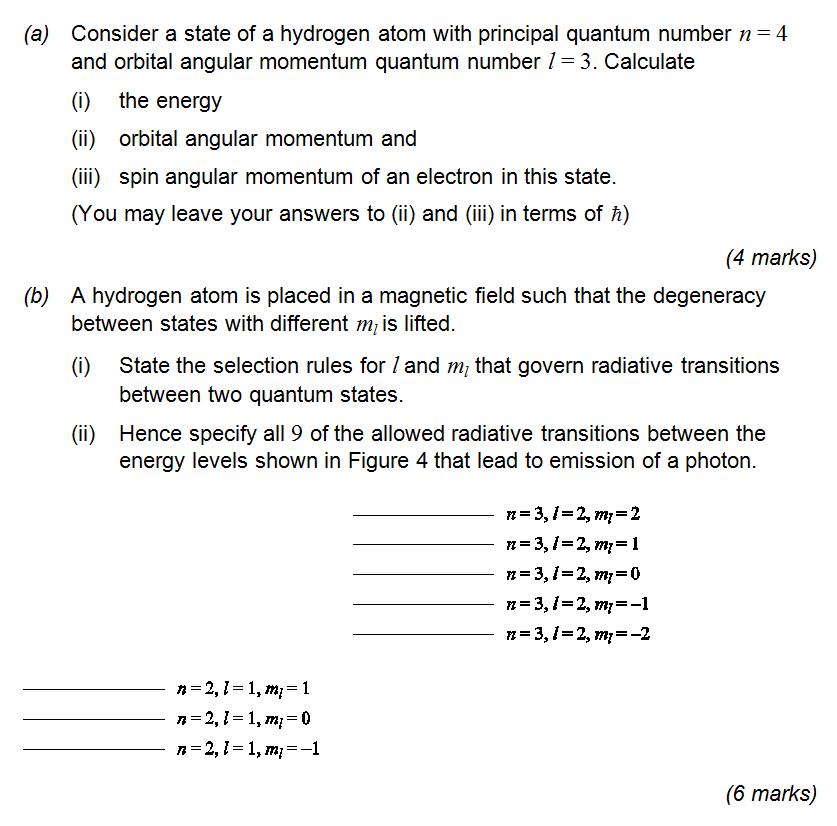}
\caption{\label{fig:14J25} Low scaffolding: 14J-25}
\end{figure}

\begin{figure}[h]
\includegraphics[scale = .57, angle = 90]{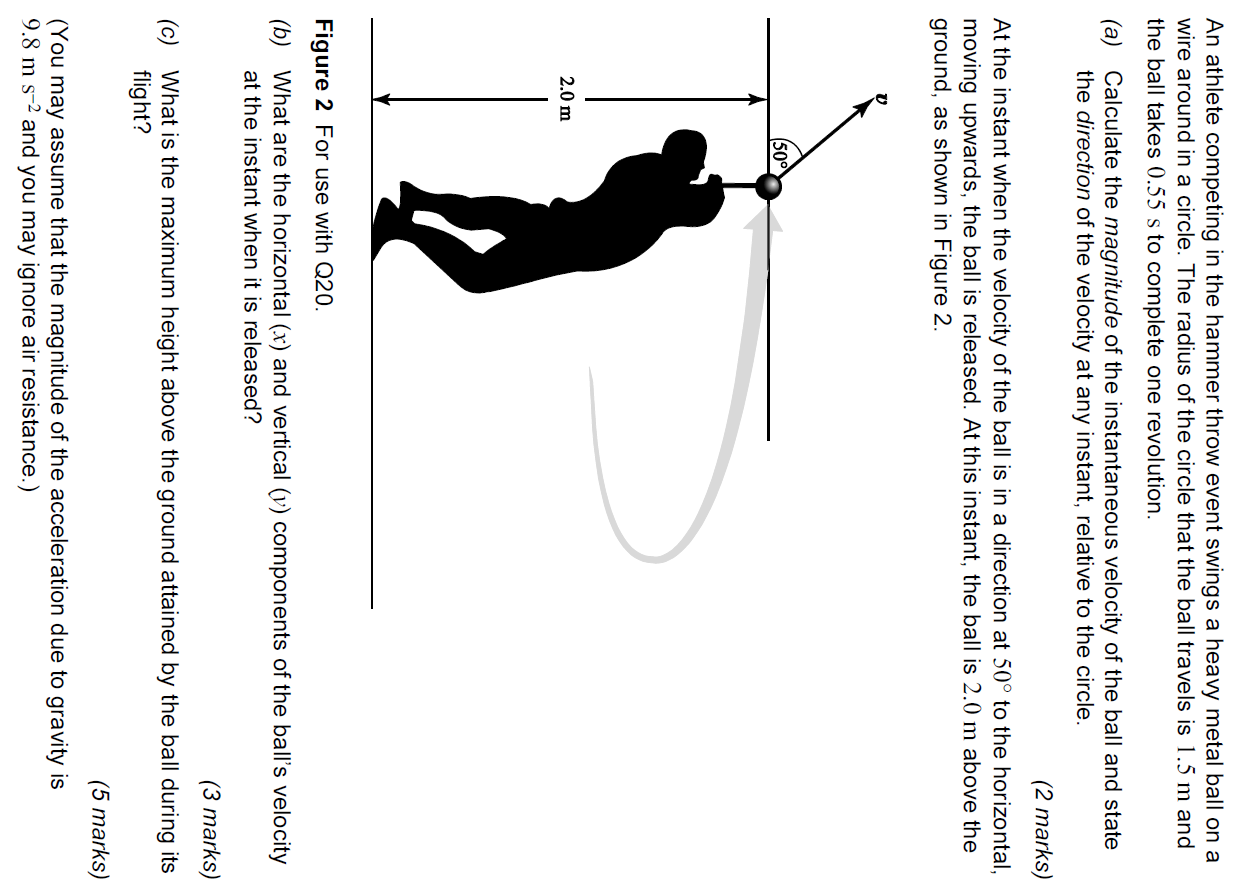}
\caption{\label{fig:14J20} High scaffolding: 14J-20}
\end{figure}

\begin{figure}[h]
\includegraphics[scale=.63]{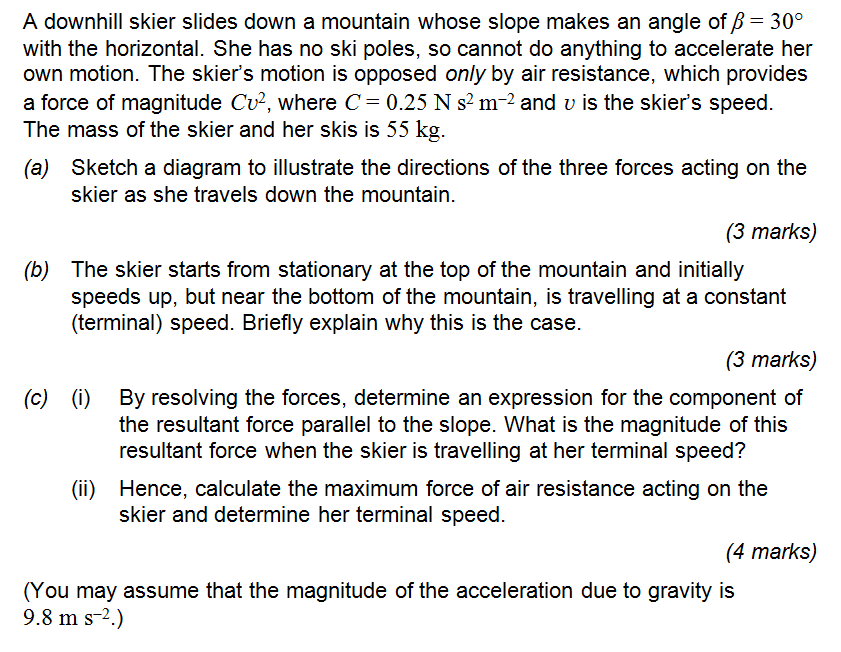}
\caption{\label{fig:14J21} High scaffolding: 14J-21}
\end{figure}

\begin{figure}[h]
\includegraphics[scale=.63]{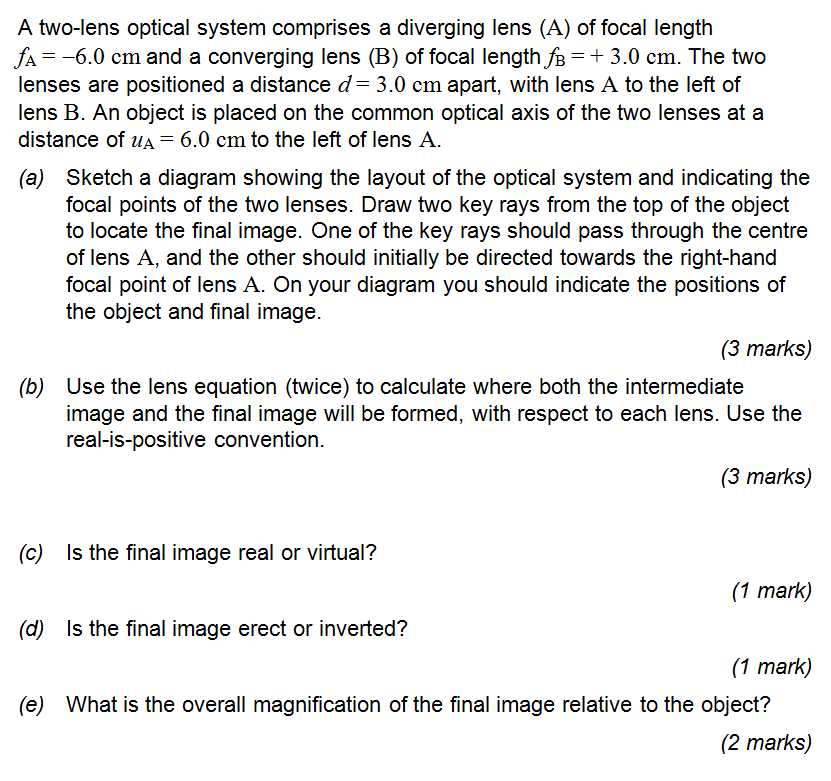}
\caption{\label{fig:14J24} High scaffolding: 14J-24}
\end{figure}

\begin{figure}[h]
\includegraphics[scale=.63]{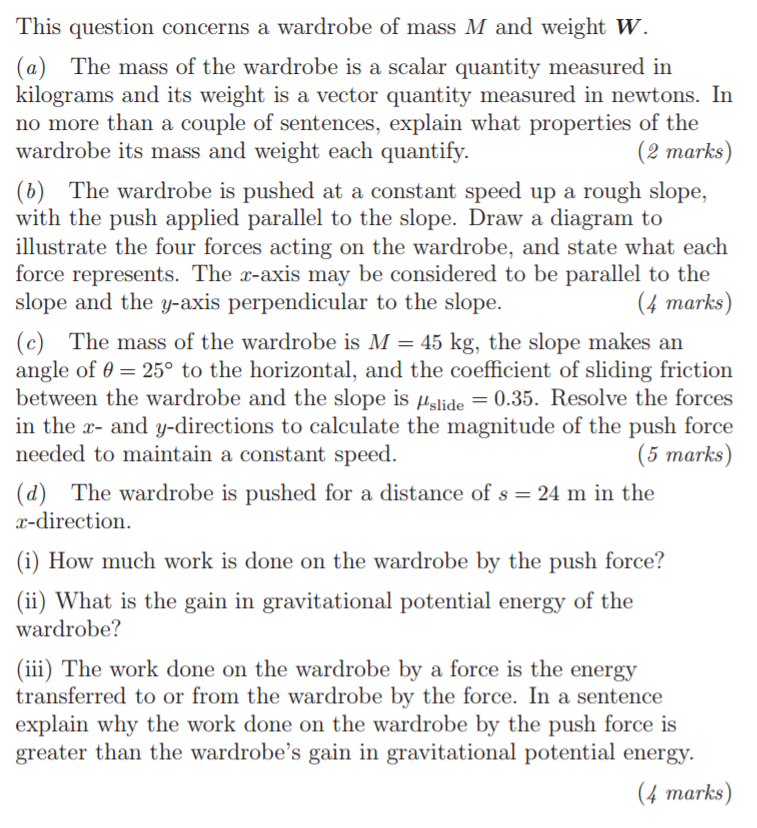}
\caption{\label{fig:15J11} High scaffolding: 15J-11}
\end{figure}

\begin{figure}[h]
\includegraphics[scale=.63]{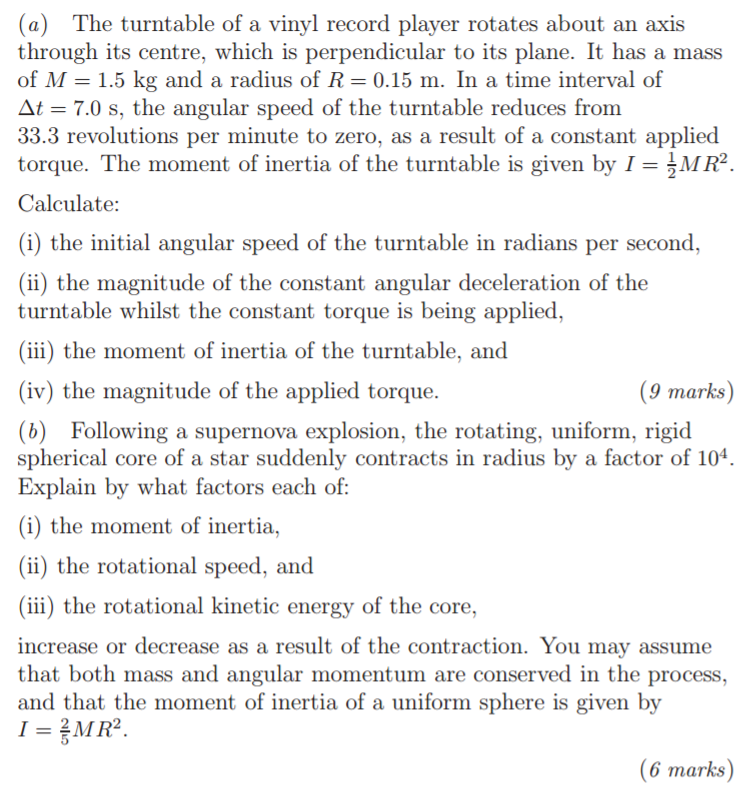}
\caption{\label{fig:15J12} High scaffolding: 15J-12}
\end{figure}

\begin{figure}[h]
\includegraphics[scale=.63]{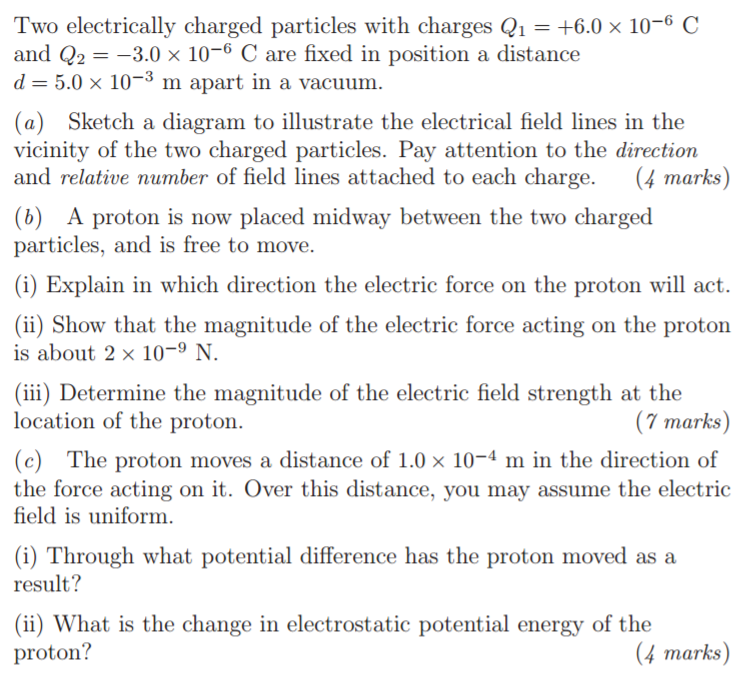}
\caption{\label{fig:15J13} High scaffolding: 15J-13}
\end{figure}

\begin{figure}[h]
\includegraphics[scale=.9]{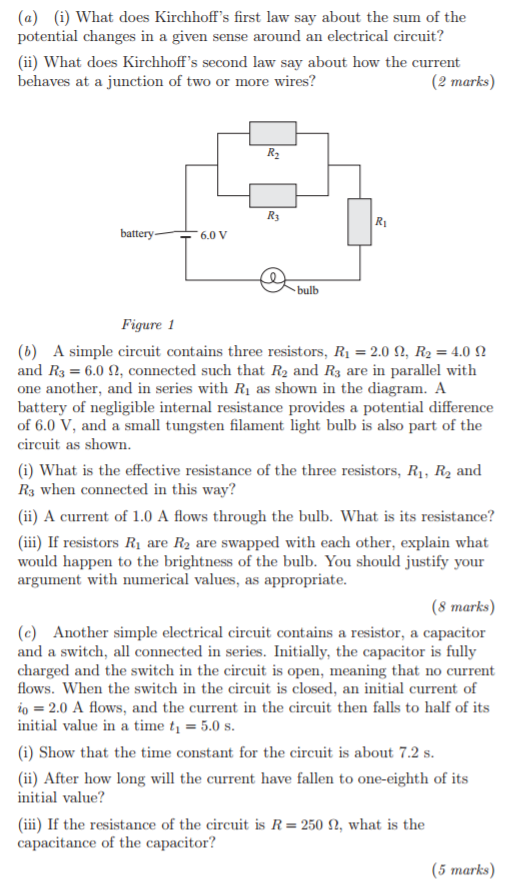}
\caption{\label{fig:15J14} High scaffolding: 15J-14}
\end{figure}

\end{document}